\documentclass[12pt]{article}
\usepackage{graphicx}
\begin{document}
\textheight=23cm
\textwidth=17.5cm
\topmargin=-1.5cm
\hoffset=-1.5cm

\centerline{\large \bf Photometric observations of}

\centerline{\large \bf Grid Giant Star Survey candidates}
\bigskip

{\bf
\noindent Bizyaev D. (UTEP), Smith V. V. (NOAO), Arenas J., \\
Geisler D. (Univ. de Concepcion, Chile)
}

\bigskip

We compare variability of radial velocity (RV) and brightness for a sample
of red giants which were selected as candidates for the Grid Giant Star
Survey (GGSS). The purpose of the survey is to select astrometrically stable
stars as a reference frame for the future Space Interferometry Mission. We
incorporate photometric data taken by ROTSE-I and being taken by ROTSE-III
robotic telescopes. Typically 2/3 of all GGSS stars to the north of DEC=-35
were observed 76 and 8 times by ROTSE-I and III respectively in white-light
filters. It is shown that photometric stability at the 0.02 mag level
cannot be used for a pre-selection of RV-stable stars, and the most
important tool remains high-resolution spectroscopy. We deeply thank the
ROTSE collaboration providing us observing time at the ROTSE-IIIb telescope.
We gratefully acknowledge NASA grant NRA-99-04-OSS-058.

\vskip 1cm

\bigskip
OBSERVATIONS
\bigskip

We present results from the spectroscopic and photometric followup survey of
GGSS candidates. The main goal of the survey is to select astrometrically
stable stars in support of the future Space Interferometry Mission. The GGSS
sample consists of mostly distant red giants (see [1]). Followup
spectroscopic observations of this sample were conducted in 2001-2004 and
reveal radial velocity stable and unstable stars at the level of 50-100 m/s. 
In support of the spectroscopic observations, we carry out a photometric
monitoring of the candidates brightness with the help of ROTSE-IIIb robotic
telescope (operated at McDonald observatory, Texas). The ROTSE-IIIb is a
fully robotic telescope ([2]) dedicated to observe GRB afterglows and well
suited for sky surveys.

Of the all-sky GGSS sample (4739 stars) we were able to observe 3233 in the
white-light filter. The expected accuracy was about 0.02 mag for our
V=11-13.5 mag stars. Since August 2003 when the survey started, we took
28401 individual magnitudes for 3233 stars. We here examine any correlation
between RV and photometric stability of our candidates.

In addition to the ROTSE-IIIb observations, we incorporate results of the
Northern Sky Variability Survey (NSVS, [3]), a white-light photometric
followup survey of all northern (DEC $>$ -35 edgrees) sky sources carried out
with the help of ROTSE-I robotic telescope (Robotic Optical Transient Search
Experiment). Approximately 2/3 our GGSS northern and southern
candidates were observed during this survey. A fraction of these objects were
rejected because of nearby companions, since the ROTSE-I pixel size is 
14 arcsec. The expected accuracy in brightness for our sample of NSVS
for our candidates is about 0.02 magnitudes. During this survey our
candidates were observed 76 times on average during one year.

\bigskip
RESULTS
\bigskip

We estimate the accuracy in brightness taking the same source 3 times per
night. The mean value of the brightness accuracy is about 0.02 mag. The
long-term brightness variability was estimated as standard deviation of all
magnitude values available for the source (sigma hereafter). Fig.1 shows the
histogram with distribution of sigma and averaged internal accuracy for 3132
stars. Dashed line shows the same values for the stars which were taken more
than two and 5 times. There is no difference in shape of these two
distributions. It is seen that the variability of brightness is
statistically significant for most of the investigated stars relative to the
internal accuracy. However, long-term night-to-night errors coming from
calibration procedure and sky conditions were not taken into account in this
error estimation. This introduces an additional error and moves the maximum
of distribution on Fig.1 along x-axis toward the higher values. It is a
reasonable assumption that most of our stars are photometrically stable.
Then the position of the maximum in Fig.1 at about 0.025 mag reveals the
real mean level of long-term photometric accuracy. From this point of view,
only a small fraction of observed GGSS candidates are photometrically
unstable.

\begin{figure}
\includegraphics[width=14cm,angle=0]{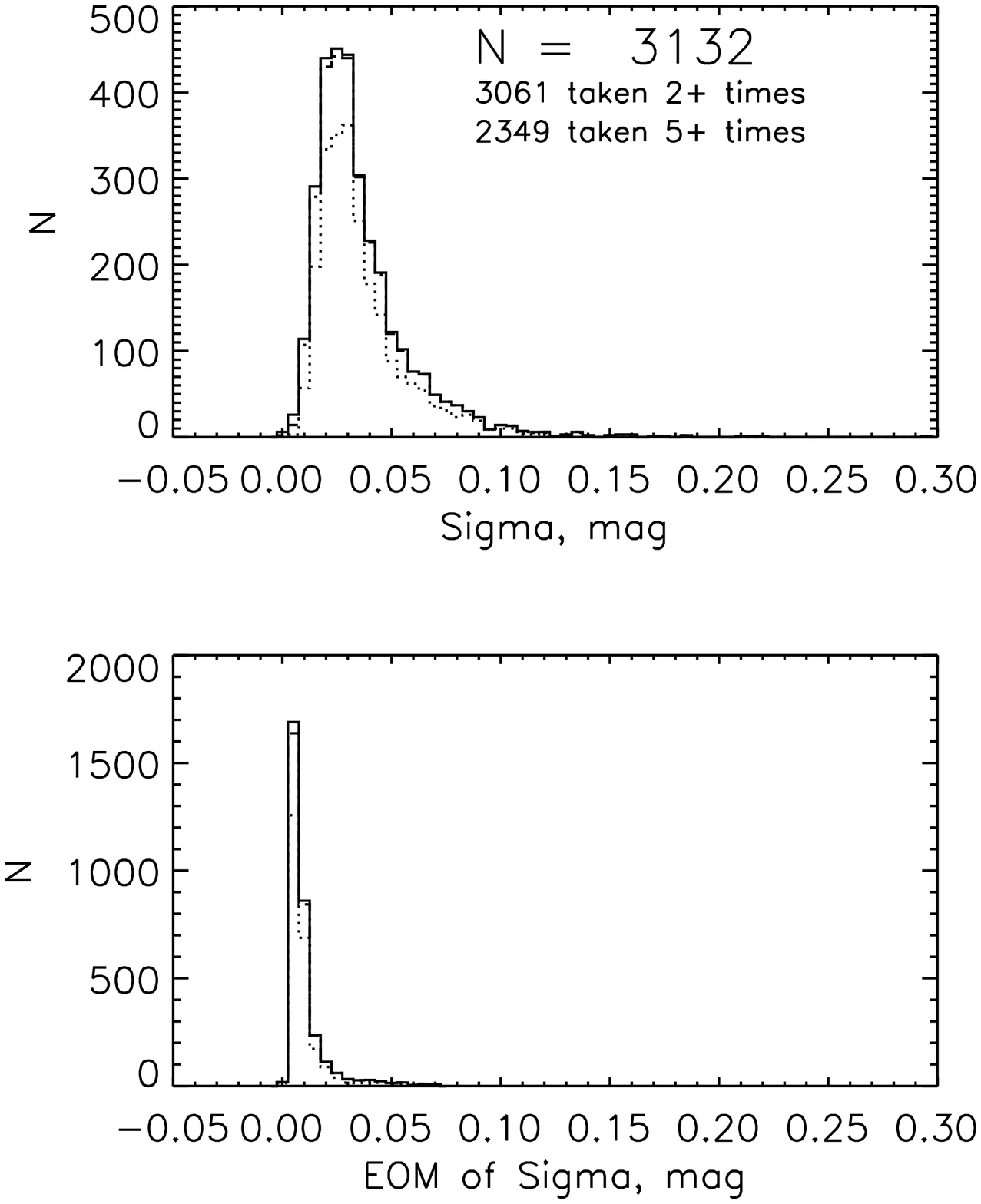}
\caption{ }
\end{figure}

We are enabled to compare the photometric variability with stability level
of radial velocity for two subsample of stars ([4]) observed at the McDonald
observatory (northern subsample hereafter) with 2.1m telescope and Sandiford
Echelle Spectrograph (R=55000), and at the ESO (Chile) with CORALIE
velocimeter at the Swiss 1.2 telescope (southern subsample). The photometric
and RV sigmas for 124 GGSS stars are compared in Fig.2 (upper left panel). 
A lack of
correlation between the photometric variability and the main stellar
parameters (Teff, log G, [Fe/H]) is seen in the rest of the panels in Fig.2.

\begin{figure}
\includegraphics[width=14cm,angle=0]{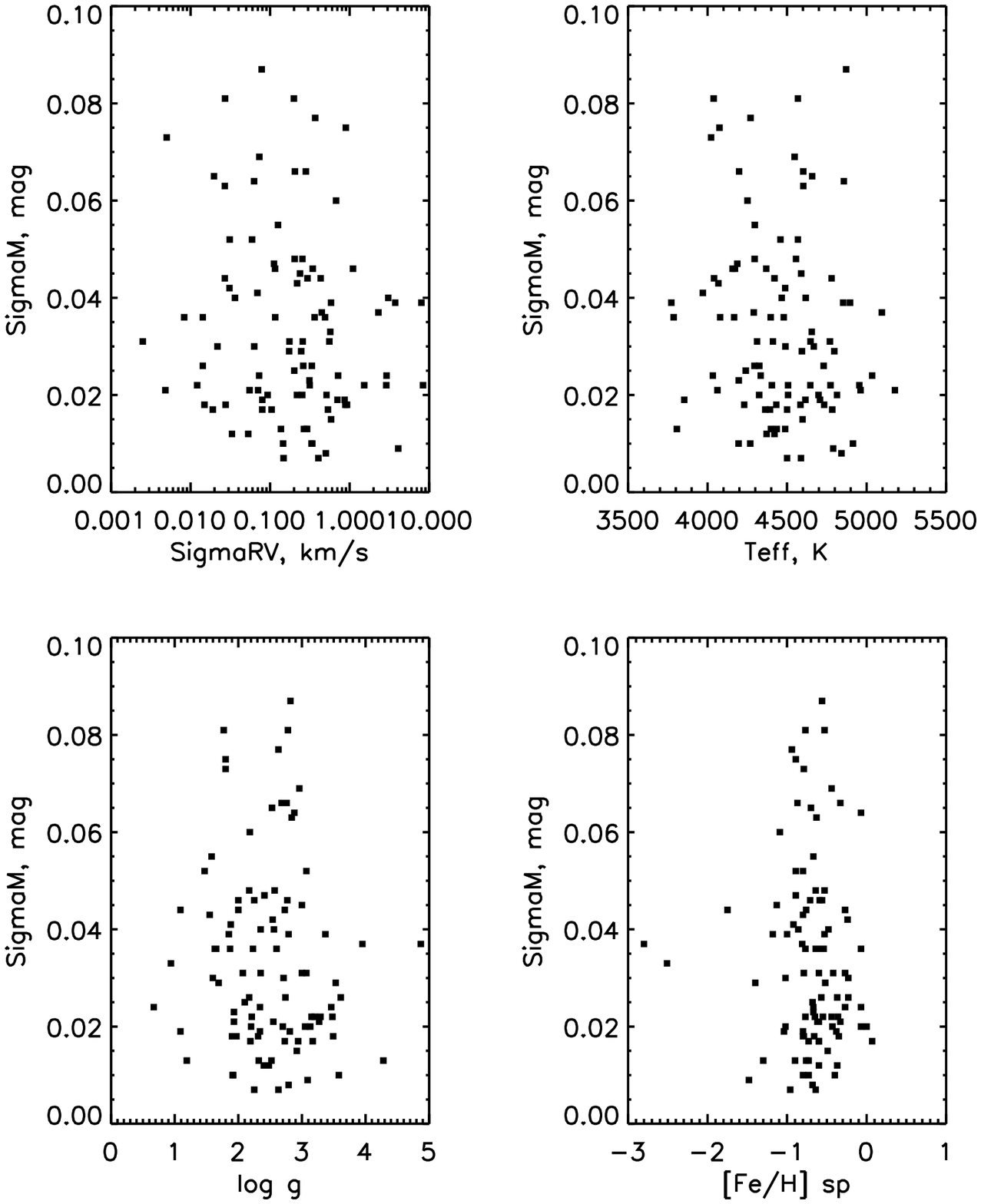}
\caption{ }
\end{figure}

The ROTSE-I photometic variability estimated as the standard deviation of
brightness measured on different nights (sigma hereafter) is also available
to compare with two samples of RV sigmas, northern and southern. Fig.3
represents the relations between photometric and RV sigmas. Left pannels
show the northern subsample, and the right ones are for the southern stars.
In addition to the photometric sigma (shown in the bottom panels), we
calculate also chi square and compare them in the top panels. Open symbols
mark the RV sigmas based on two estimates, and the filled ones are for 3 and
more RV values utilized to obtain the RV sigma.

\begin{figure}
\includegraphics[width=14cm,angle=0]{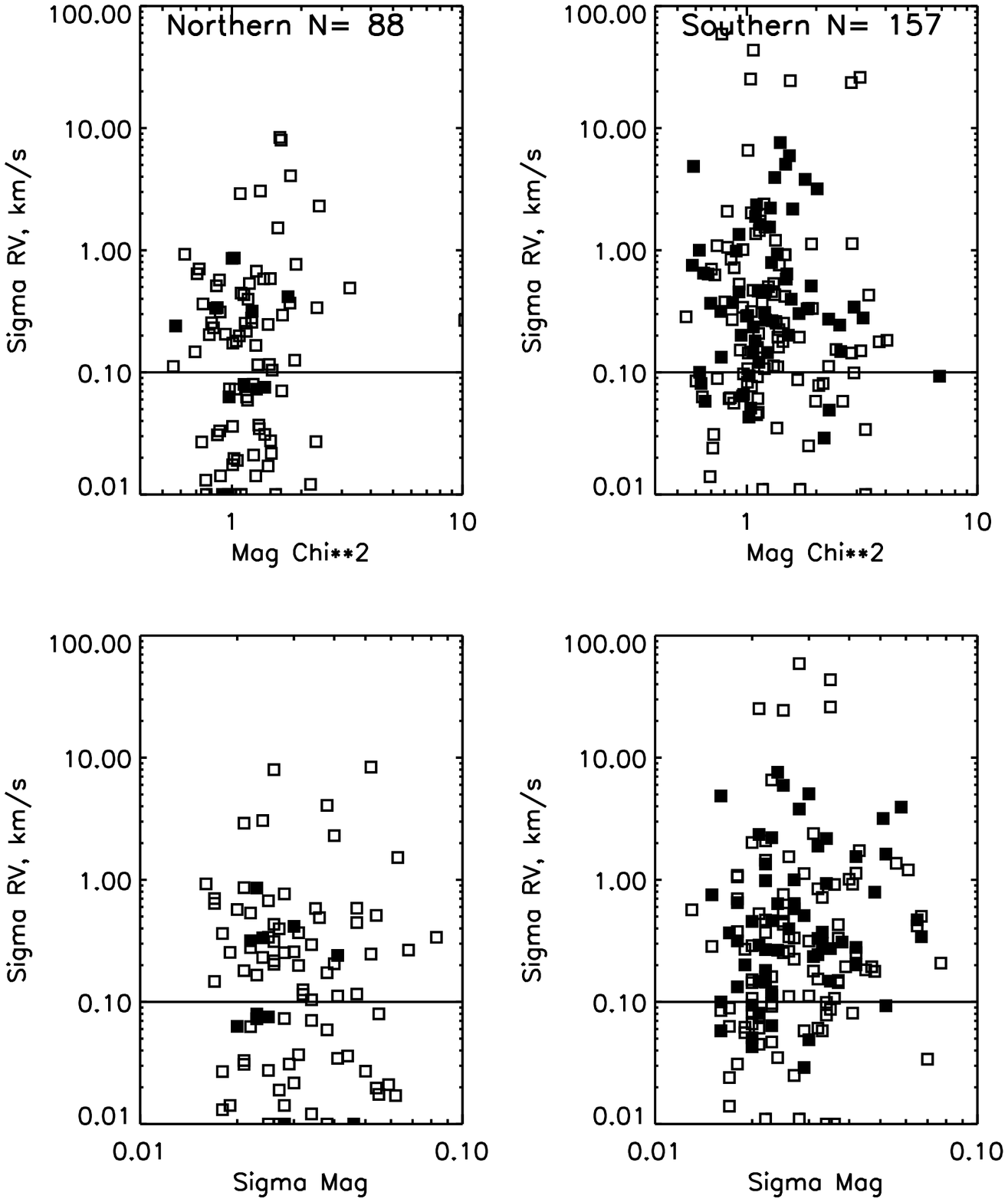}
\caption{ }
\end{figure}

There is a lack of correlation between the basic stellar parameters and chi 
square photometric values, as well as for the photometric sigma, obtained for 
the ROTSE-I data. Finally, we kept an eye on individual objects and found 
some examples of stars with highly variable radial velocity but stable 
photometrically, as well as vice versa, good and stable radial velocity 
can coexist with suspicious photometric instability.

\bigskip
CONCLUSION
\bigskip

We investigate whether the all-sky photometric variability surveys can help
us to pre-select K-giants wich would indicate stable radial velocity.
We show that, most probably, it can not be done via a photometric survey,
at least with accuracy of order 0.025 mag. The most important tool remains
high-resolution spectroscopy.

\bigskip
ACKNOWLEDGMENTS
\bigskip

We deeply thank the ROTSE collaboration, and personally Carl Akerlof, Tim
McKey, Eli Rykoff and Don Smith for providing us observing time at the
ROTSE-IIIb telescope and for discussions.  We thank Michel Mayor for
generous allotment of telescope time.  This project utilizes data obtained
by the Robotic Optical Transient Search Experiment.  ROTSE is a
collaboration of Lawrence Livermore National Lab, Los Alamos National Lab,
and the University of Michigan. The GGSS followup survey is supported by
NASA/JPL via grant NRA-99-04-OSS-058.

\bigskip
REFERENCES
\bigskip

{\small
[1] Patterson R., Majewski S., Kundu A. et al., 1999, AAS 195, 4603

[2] Akerlof C., Kehoe R., McKay T. et al., 2003, PASP 115, 132

[3] Wozniak P., Westrand W., Akerlof C. et al., 2004, AJ 127, 2436

[4] Bizyaev D., Smith V. V., Arenas J.et al., 2003, AAS 203, 4904

}

\end{document}